\documentclass{article}
\usepackage{amsmath,graphicx,mlspconf}
\usepackage{xcolor}
\usepackage{comment}
\usepackage{multirow}
\usepackage{url}
\usepackage{booktabs}

\copyrightnotice{U.S.\ Government work not protected by U.S.\ copyright}

\copyrightnotice{979-8-3503-2411-2/25/\$31.00 {\copyright}2025 Crown}

\copyrightnotice{979-8-3503-2411-2/25/\$31.00 {\copyright}2025 European Union}

\copyrightnotice{979-8-3195-0884-3/26/\$31.00 {\copyright}2026 IEEE}

\toappear{2026 IEEE International Workshop on Machine Learning for Signal Processing, Sep.\ 28-- Oct.\ 1, 2026, Atlanta, USA}

\title{Mitigating Over-Suppression in Speech Enhancement via Inference-Time Rethink-and-Refine Correction Module}
\name{Mike Qu$^{*}$, Yu-Wen Chen$^{*}$, Julia Hirschberg\thanks{$^{*}$Equal contribution.}}

\address{
    Department of Computer Science, Columbia University, USA
}

\begin{document}

\maketitle

\begin{abstract}
We present a rethink-and-refine correction module that addresses over-suppression, a common failure mode of speech enhancement (SE) models, where speech cues are suppressed alongside noise. Our method operates entirely in the inference stage without additional training, allowing seamless integration with diverse SE models. Given noisy and enhanced signals, we obtain word- or phoneme-level alignments using an automatic speech recognition model and identify intervals where enhancement is unreliable. These intervals are then selectively remixed through convex interpolation, with per-segment weights optimized to maximize a composite objective balancing perceptual quality and speech preservation. Experiments on the URGENT 2024 and 2025, VCTK-DEMAND, and MSP-PODCAST datasets show consistent improvements in perceptual quality, intelligibility, and downstream performance compared to conventional SE alone, demonstrating the benefit of rethink-and-refine framework for robust speech processing.
\end{abstract}

\newcommand{\cem}[1]{\textcolor{blue}{cem: #1}}
\section{Introduction}
\label{sec:intro}

Speech enhancement (SE) models aim to suppress noise while preserving speech quality and intelligibility. Recent deep learning-based approaches~\cite{Cao_2022, chao2024investigationincorporatingmambaspeech, richter2023speechenhancementdereverberationdiffusionbased} are typically trained on tuples of noisy and clean waveforms, with the objective of manipulating the noisy signal to make it as “similar” to the clean signal as possible~\cite{pandey2021, Lu_2023}. When test conditions match training data, these models often achieve substantial gains on objective and subjective assessment metrics. However, when deployed in unseen or complex real-world noise environments, SE models often produce unnatural artifacts or destructive distortions. One mitigation strategy is to train or fine-tune the model on data resembling the target environment; however, in many real-world applications, the clean and noise data required for SE training are unavailable. To address this, output-level correction offers an alternative by treating the enhancer as a black box and operating only on its output. These methods adjust the enhanced waveform without fine-tuning the base model, enabling use across different SE and noise conditions. For example, observation-adding approaches address suppression artifacts by globally interpolating the noisy and enhanced signals, relying on a fixed global mixing rule to mitigate distortions introduced during enhancement ~\cite{iwamoto2022badartifactsanalyzingimpact}. NRSER used a signal-to-noise ratio (SNR)-based detector to select the interpolation weight between SE output and the original signal, improving the noise robustness of downstream emotion recognition~\cite{chen2023noiserobustspeechemotion}.

A common failure mode for SE is under-enhancement, in which residual background noise remains after processing. Far more harmful is \emph{over-suppression}, where the model over-suppresses the signal to the extent that it substantially removes parts of the speaker’s information, leaving the output less intelligible and less informative for downstream tasks, sometimes even worse than unprocessed signals. The performance of SE is typically evaluated using speech assessment (SA) metrics. Compared to traditional intrusive metrics like PESQ (perceptual evaluation of speech quality)~\cite{rix2001pesq} and STOI (short-time objective intelligibility)~\cite{Taal2011Algorithm}, non-intrusive metrics such as mean opinion score (MOS) predictors estimate perceived listening quality without requiring clean references and are sensitive to both residual noise and over-suppression. While prior work has incorporated such predictors into training objectives \cite{pmlr-v202-shin23b, close2023multicmganleveragingmultiobjectivespeech}, leveraging SA models at test time for output-level correction, especially to prevent destructive over-suppression, remains largely unexplored.

We reformulate the output-level correction as a “rethink-and-refine” strategy that is prevalent in natural language processing. Large language models frequently produce an initial answer that is then checked, critiqued, and selectively edited using auxiliary modules or downstream signals, rather than being treated as a final output \cite{shinn2023reflexionlanguageagentsverbal, madaan2023selfrefineiterativerefinementselffeedback}. The key insight in these frameworks is that generation and evaluation can be separated: the model produces a first-pass output, and a later stage equipped with stronger or more specialized feedback modifies the parts that need improvement. This decoupling enables the incorporation of feedback signals, such as human judgments or assessment models, that would have been difficult to enforce during initial generation itself. By analogy, SE, which is typically performed in a single forward pass, may also benefit from a secondary correction stage that revisits the enhanced waveform with SA, identifies localized errors such as over-suppression, and repairs them without altering regions that were already handled correctly.

Together, these observations motivate an output-level inference-time refinement approach for SE under the guidance of automatic speech recognition (ASR) and non-intrusive SA models. Using ASR timestamps, we align noisy and enhanced audio at the word or phoneme level and optimize a mixing weight for each segment to maximize perceptual quality. Our key contributions are these: (1) The proposed module operates without any additional training and can be applied to any off-the-shelf SE model. Serving as a plug-in module, it alleviates the catastrophic over-suppression that arises when SE models are deployed under unseen acoustic conditions. (2) Our module achieves consistent gains in intrusive (e.g., PESQ and STOI) across SE models and datasets. (3) Experimental results further demonstrate improved performance on downstream tasks, with reduced word error rates (WER) of transcription and better preservation of vocal traits.

\begin{figure*}[!t]
    \vspace{-4mm}
    \centering
    \includegraphics[width=0.80\textwidth]{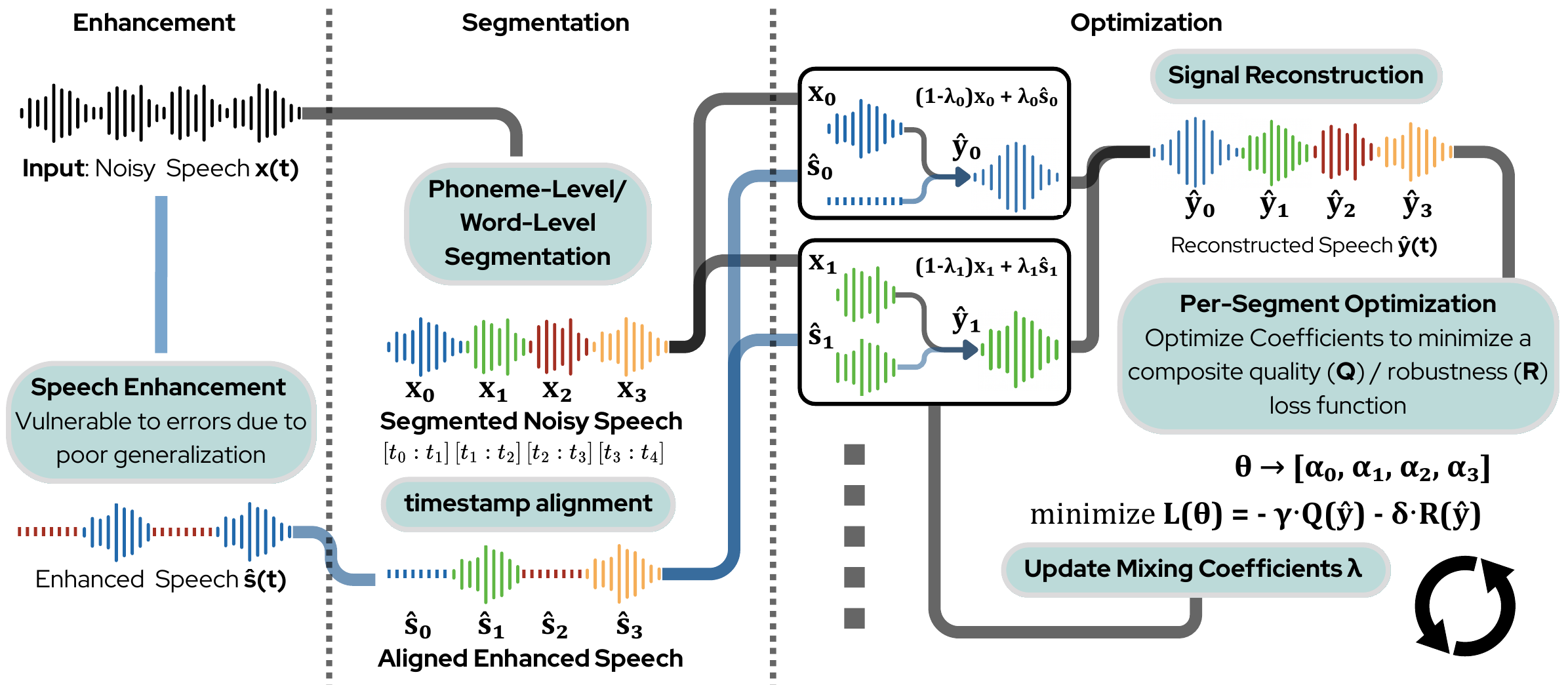}
    \caption{The proposed correction module. The module can be added to any SE model, operating directly on its output without additional training. An ASR system segments the speech signal, and a SA model guides the correction of each segment.}
    \label{fig:architecture}
    \vspace{-4mm}

\end{figure*}

\section{Methodology}
\label{sec:methods}

We propose a correction module for SE, inspired by the rethink-and-refine framework, which identifies parts of the input speech where the initial SE over-suppresses the signal and selectively restores the information (Fig.~\ref{fig:architecture}). First, we segment both the noisy and output of pretrained SE model (i.e., enhanced signals) into word- or phoneme-level intervals using ASR-based timestamps, forming paired segments that are aligned over the same time spans. For each segment, we then optimize a mixing weight that interpolates between noisy and enhanced signal, maximizing a perceptual quality and robustness objective on the reconstructed waveform with SA. In addition, a reconciliation strategy is introduced to address the issue of segment boundary mismatches. 


\subsection{ASR-guided segmentation}
To localize and repair regions affected by over-suppression, we introduce a post-processing stage based on word- or phoneme-level segmentation. Given a noisy signal $x(t)$ and its enhanced counterpart $\hat{s}(t)$, we run an ASR-based segmentation model, including Charsiu (for phonemes)~\cite{zhu2022charsiu}  or Whisper-timestamped (for words)~\cite{radford2022robust, JSSv031i07}. 

This produces a sequence of time intervals
\begin{equation*}
[t_i^{\mathrm{start}}, t_i^{\mathrm{end}}], \quad i = 1,\dots,K
\end{equation*}
where each interval corresponds to a meaningful speech unit.

For each interval, we extract the corresponding segments from noisy $x_i(t)$ and enhanced $\hat{s}_i(t)$ signals. These segments serve as the basic units for the subsequent mixing process. The interval-wise view of noisy segment and enhanced segment makes it easier to detect over-suppression (i.e., speech present in $x_i$ but missing or strongly attenuated in $\hat{s}_i$) than working with the full waveform, where such differences can be obscured by small timing mismatches.




\subsection{SA-guided segment-wise weight selection}
Over-suppression occurs when the SE model removes genuine speech energy along with noise, where the noisy segment $x_i(t)$ contains speech cues that have been suppressed in enhanced signal $\hat{s}_i(t)$, while $\hat{s}_i(t)$ may still provide useful denoising. Rather than discarding either source, we repair these regions by interpolation, with weights guided by non-intrusive SA metrics. Specifically,

For each segment, we define a convex interpolation
\begin{equation} \label{eqn:mixing}
    s_i^{\mathrm{corr}}(t, \lambda_i) 
    = \lambda_i \,\hat{s}_i(t) + (1-\lambda_i)\,x_i(t),
    \qquad \lambda_i \in [0,1],
\end{equation}

where $\lambda_i$ is a segment-wise mixing weight: values near $1$ favor the enhanced signal, and values near $0$ favor the original noisy input. Let $\lambda = [\lambda_1,\dots,\lambda_K]$ denote the vector of mixing weights over all segments. A corrected waveform $s^{\mathrm{corr}}(t,\lambda)$ is obtained by replacing each segment in time with $s_i^{\mathrm{corr}}(t,\lambda_i)$ and leaving the rest of the signal unchanged.


To select $\lambda$, we optimize a composite objective that balances perceptual quality and robustness on the reconstructed signal. Let $Q(\cdot)$ denote a non-intrusive quality predictor (e.g.,  SCOREQ), and let $R(\cdot)$ measure the preservation of speech-like structure (e.g., Whisper embedding L2 similarity). We define
\begin{equation}
    \mathcal{L}(\lambda) 
    = \gamma\, Q\big(s^{\mathrm{corr}}(t,\lambda)\big) 
    + \delta\, R\big(s^{\mathrm{corr}}(t,\lambda)\big),
\end{equation}

with weights $\gamma$ and $\delta$ controlling the trade-off (both set to $0.5$ in our experiments). A higher $\gamma$ biases the optimizer to focus on the raw quality of the produced speech signal, while a higher $\delta$ places a heavier emphasis on the preservation of speech cues and semantics. The optimal mixing vector is then
\begin{equation} \label{eqn:lambda}
    \lambda^\star = \arg\max_{\lambda \in [0,1]^K} \mathcal{L}(\lambda).
\end{equation}

This formulation treats the segment weights as inference-time parameters, enabling adaptation of any given SE output without retraining or modifying the base model. In practice, $\lambda^\star$ is obtained via continuous gradient descent with the Adam optimizer, initialized at $\lambda=0.5$. Existing methods~\cite{iwamoto2022badartifactsanalyzingimpact, chen2023noiserobustspeechemotion} employ a similar formulation but restrict the mixing coefficient to be a \emph{single} global value or a coarsely selected utterance-level quantity (e.g., based on an SNR proxy), implicitly assuming a uniform correction strength across time \cite{iwamoto2022badartifactsanalyzingimpact, chen2023noiserobustspeechemotion}. Our method further generalizes that approach to ASR-aligned segments and treats the set of mixing weights $\{\lambda_i\}_{i=1}^K$ as inference-time decision variables, selected by directly optimizing the composite objective in Equation \ref{eqn:lambda}. This enables content-dependent correction that selectively addresses over-suppression artifacts while avoiding unnecessary reintroduction of noise in well-enhanced regions.


\subsection{Reconciliation}
\label{sec:reconciliation}
The noisy-based and enhanced-based segments generally do not share identical boundaries, so attempting to merge them at the interval level would require explicit boundary matching or temporal warping, which is brittle in practice. Therefore, we adopt a simple reconciliation strategy instead. We consider two complementary segmentation views: one obtained from the noisy signal and one from the enhanced signal. In both cases, the construction above yields aligned pairs $(x_i, \hat{s}_i)$ cut over the same time spans. We then run the correction procedure described above twice, once using the intervals obtained from the noisy signal and once using the intervals obtained from the enhanced signal. This yields two fully reconstructed candidates, each consistent with its own segmentation view. We then evaluate both candidates with the same objective $\mathcal{L}(\cdot)$ and select the higher-scoring one as the final corrected signal. In this way, both views can contribute: one emphasizing preservation of speech activity, the other emphasizing regions already improved by the SE model, without requiring explicit alignment between their segment boundaries.

To ensure that the correction process does not degrade quality in cases where remixing is unnecessary or counterproductive, we additionally compare the selected reconstruction against the original noisy and enhanced signals under the same objective. If either baseline achieves a higher score, we retain it instead. This fallback criterion ensures that the correction stage is conservative and never performs worse than simply choosing the better of the two original inputs.

\section{Experimental setup}

\subsection{Models and datasets}
We evaluate our approach using three state-of-the-art SE models, including CMGAN~\cite{Cao_2022}, SEMamba~\cite{chao2024investigationincorporatingmambaspeech}, and SGMSE~\cite{richter2023speechenhancementdereverberationdiffusionbased}. For fair comparison, all models use official VoiceBank-DEMAND pre-trained checkpoints: the standard CMGAN checkpoint, \texttt{SEMamba\_advanced.pth}, and SGMSE+. To guide weight selection, we use Whisper-base~\cite{radford2022robust} as the ASR embedding model used to compute a L2 distance-based robustness objective, and SCOREQ~\cite{ragano2025scoreqspeechqualityassessment} as the SA model used to compute the speech quality objective.

We conducted experiments on a collection of publicly available speech datasets that span diverse speakers, languages, acoustic conditions, and noise environments. This includes the non-blind test sets of the URGENT 2024 \cite{Zhang_2024} and 2025 \cite{saijo2025interspeech2025urgentspeech} challenges, the VCTK-DEMAND dataset \cite{valentinibotinhao2016investigating}, as well as MSP-PODCAST~\cite{lotfian2017building} mixed with Audioset~\cite{gemmeke2017audioset} noise at two SNR levels (4 and 8), as settings in~\cite{chen2023noiserobustspeechemotion}. These benchmarks provide distinct and complementary evaluation conditions: URGENT 2024 contains English speech with diverse noise profiles, while URGENT 2025 extends evaluation to multilingual settings. VCTK-DEMAND represents lower-intensity stationary noise conditions, and MSP-PODCAST/Audioset enables testing with large-scale noise sources at configurable SNRs. Together, they cover a broad range of test-time conditions, allowing robust evaluation under a variety of potentially unseen scenarios.

\subsection{Evaluation metrics and models}

We evaluate our method using both SA metrics and measures that reflect downstream tasks. For SA, we report standard intrusive metrics, including PESQ and STOI, alongside non-intrusive quality scores from the Meta Audiobox aesthetics evaluator (AB-Aes.), specifically the content enjoyment (CE) and perceptual quality (PQ) axes~\cite{tjandra2025metaaudioboxaestheticsunified}. We also include the SCOREQ MOS score~\cite{ragano2025scoreqspeechqualityassessment}, which was used to guide the weight selection. For downstream tasks, we evaluate the WER of an ASR model (i.e., Whisper-base~\cite{radford2022robust}), calculated as the edit distance between transcripts from clean speech and those generated from noisy, initial SE, and our method. We also measure the preservation of speaker characteristics. We evaluate the speaker verification score, computed using the $L_2$ distance of NVIDIA’s SpeakerNet embeddings~\cite{koluguri2020speakernet1ddepthwiseseparable} between the clean reference and the corresponding noisy, initial SE, and our method. Additionally, jitter and shimmer, which capture characteristics such as pathological conditions~\cite{teixeira2013vocal} or the speaker’s emotional state~\cite{bone2014psychologist}, are calculated using the Parselmouth library~\cite{parselmouth}\footnote{\url{http://www.praat.org/}}. Subjective listening tests were conducted on the URGENT 2024 dataset, chosen for its English-only speech under diverse and challenging noise conditions. Nine annotators were divided into three groups, each assigned to evaluate a set of audio samples comprising a balanced mixture of clean speech, noisy speech, initial SE outputs, and our corrected outputs. In total, 240 audio samples were evaluated across all groups. Samples were rated on a 1–5 scale (bad to excellent), where 1 indicates severe distortion with unintelligible words and 5 denotes recording-quality audio. Annotation was performed using Label Studio~\footnote{\url{https://labelstud.io}}.

\section{Results}
\subsection{Performance across datasets}

We evaluate SEMamba~\cite{chao2024investigationincorporatingmambaspeech} both with and without the proposed output-level correction module across multiple benchmark datasets to examine whether our approach can reliably improve a strong SE under unseen noise conditions (Table~\ref{tab:main-results}). On every dataset, our method yields higher STOI and SCOREQ MOS than the enhanced baseline, and also improves PESQ and Meta Audiobox aesthetic scores (CE/PQ) in most cases. The relatively lower performance on the VCTK DEMAND dataset may be attributed to its higher average SNR as well as the fact that SEMamba was trained on DEMAND noise, which reduces the occurrence of over-suppression. Consequently, the original SEMamba already performs strongly, leaving limited room for our correction module to provide additional benefits. Overall, the gains are stable across recording conditions and SNR levels, indicating that the proposed method reliably adds measurable perceptual improvements on top of a pretrained SE model, even when the base SE already performs strongly.

\begin{table}[t]
\centering
\caption{Performance across datasets (mean {\scriptsize$\pm$} std).}
\label{tab:main-results}
\setlength{\tabcolsep}{2pt}
\renewcommand{\arraystretch}{1.00}
\resizebox{0.87\linewidth}{!}
{%
\begin{tabular}{llccccc}
\toprule
\textbf{Dataset} & \textbf{Cond.} & PESQ & STOI & AB-Aes & AB-Aes & SCOREQ \\
-- & -- & $\uparrow$ & $\uparrow$ & CE$\uparrow$ & PQ$\uparrow$ & MOS$\uparrow$ \\
\midrule
\multirow{4}{*}{\begin{tabular}{l}URGENT\\2025\end{tabular}}
& clean & 4.644 & 1.000 & 4.851 & 6.172 & 3.321 \\
& noisy & 1.306 & 0.753 & 3.751 & 4.921 & 1.763 \\
\cmidrule(lr){2-7}
& enh. & 1.535 & 0.678 & 3.600 & 5.109 & 2.044 \\
& ours & \textbf{1.546} & \textbf{0.741} & \textbf{3.755} & \textbf{5.142} & \textbf{2.221} \\
\midrule
\multirow{4}{*}{\begin{tabular}{l}URGENT\\2024\end{tabular}}
& clean & 4.644 & 1.000 & 5.393 & 6.352 & 3.978 \\
& noisy & 1.415 & 0.840 & 3.994 & 4.746 & 2.252 \\
\cmidrule(lr){2-7}
& enh. & 1.941 & 0.833 & 4.453 & 5.399 & 3.055 \\
& ours & \textbf{1.961} & \textbf{0.859} & \textbf{4.495} & \textbf{5.358} & \textbf{3.168} \\
\midrule
\multirow{4}{*}{\begin{tabular}{l}VCTK\\DEMAND\end{tabular}}
& clean & 4.644 & 1.000 & 5.664 & 6.352 & 3.978 \\
& noisy & 1.970 & 0.921 & 4.694 & 4.746 & 2.252 \\
\cmidrule(lr){2-7}
& enh. & \textbf{3.559} & 0.960 & 4.453 & \textbf{5.399} & 3.055 \\
& ours & 3.415 & \textbf{0.961} & \textbf{4.495} & 5.358 & \textbf{3.168} \\
\midrule
\multirow{4}{*}{\begin{tabular}{l}MSP\\SNR-4\end{tabular}}
& clean & 4.644 & 1.000 & 5.279 & 6.263 & 3.751 \\
& noisy & 1.238 & 0.804 & 4.396 & 5.326 & 2.024 \\
\cmidrule(lr){2-7}
& enh. & 1.570 & 0.830 & 4.567 & 5.901 & 3.061 \\
& ours & \textbf{1.585} & \textbf{0.842} & \textbf{4.594} & \textbf{5.904} & \textbf{3.150} \\
\midrule
\multirow{4}{*}{\begin{tabular}{l}MSP\\SNR-8\end{tabular}}
& clean & 4.644 & 1.000 & 5.279 & 6.263 & 3.751 \\
& noisy & 1.394 & 0.862 & 4.541 & 5.379 & 2.308 \\
\cmidrule(lr){2-7}
& enh. & 1.799 & 0.877 & 4.831 & \textbf{6.126} & 3.348 \\
& ours & \textbf{1.840} & \textbf{0.888} & \textbf{4.834} & 6.101 & \textbf{3.430} \\
\bottomrule
\end{tabular}%
} 
\end{table}

\subsection{Performance across SE models}\

To evaluate the generalizability of our module, we integrated it with three distinct SE models (SEMamba, CMGAN, and SGMSE). As shown in Table~\ref{tab:enhancement_models}, our method yields consistent improvements over each model’s initial enhanced output. For all three enhancers, STOI and PESQ increase after applying our method. The Audiobox aesthetic scores (CE/PQ) also improve in most cases, with only small variations for SGMSE where the baseline CE and PQ scores are already high. These results indicate that the proposed output-level correction provides measurable perceptual gains across diverse SE architectures without requiring model-specific tuning.

\begin{table}[htbp!]
\centering
\caption{Performance across SE models (mean {\scriptsize$\pm$} std).}
\label{tab:enhancement_models}
\setlength{\tabcolsep}{2pt}
\renewcommand{\arraystretch}{1.00}
\scalebox{0.87}{%
\begin{tabular}{llccccc}
\toprule
\textbf{Enhancer} & Cond. & PESQ & STOI & AB-Aes & AB-Aes & SCOREQ \\
-- & -- & $\uparrow$ & $\uparrow$ & CE$\uparrow$ & PQ$\uparrow$ & MOS$\uparrow$ \\
\midrule
\multirow{2}{*}{Reference}
& clean & 4.644 & 1.000 & 4.851 & 6.172 & 3.321 \\
& noisy & 1.306 & 0.753 & 3.751 & 4.921 & 1.763 \\
\midrule
\multirow{2}{*}{CMGAN}
& enh. & 1.537 & 0.682 & 3.566 & 5.048 & 2.029 \\
& ours & \textbf{1.558} & \textbf{0.759} & \textbf{3.774} & \textbf{5.087} & \textbf{2.155} \\
\midrule
\multirow{2}{*}{SEMamba}
& enh. & 1.535 & 0.678 & 3.600 & 5.109 & 2.044\\
& ours & \textbf{1.546} & \textbf{0.741} & \textbf{3.755} & \textbf{5.142} & \textbf{2.221} \\
\midrule
\multirow{2}{*}{SGMSE}
& enh. & 1.423 & 0.658 & \textbf{4.084} & \textbf{5.716} & 2.370 \\
& ours & \textbf{1.464} & \textbf{0.713} & 4.050 & 5.554 & \textbf{2.406} \\

\bottomrule
\end{tabular}%
} 
\end{table}

\subsection{Performance on downstream tasks}
Table~\ref{tab:downstream-tasks} shows the evaluation on downstream tasks. Compared with the SEMamba outputs with and without our proposed module, our approach consistently lowers WER and SpkVer relative to the enhanced baseline across all datasets, indicating fewer ASR and speaker-embedding distortions after correction. Jitter and shimmer remain similar, with small improvements or trade-offs depending on the dataset. We observe that noisy speech yields better results than enhanced speech, possibly because ASR and speaker verification models may already incorporate inherent noise-robust mechanisms, which could conflict with additional SE. However, as shown earlier, enhanced speech generally achieves higher PESQ and STOI scores than noisy speech. Similar observations have also been reported in~\cite{de2026too}, suggesting that model-based scores may obscure the effects of residual noise.

\begin{table}[t]
\centering
\caption{Performance on downstream tasks (mean {\scriptsize$\pm$} std).}
\label{tab:downstream-tasks}
\setlength{\tabcolsep}{2pt}
\renewcommand{\arraystretch}{1.00}
\resizebox{0.87\linewidth}{!}
{%
\begin{tabular}{llcccc}
\toprule
\textbf{Dataset} & \textbf{Cond.} & WER$\downarrow$ & SpkVer$\downarrow$ & Jitter$\leftrightarrow$ & Shimmer$\leftrightarrow$\\
-- & -- & ($\%$) & -- & ($\times10^{-2}$) & ($\times10^{-2}$) \\
\midrule
\multirow{4}{*}{\begin{tabular}{l}URGENT\\2025\end{tabular}}
& clean    & 0.0  & 0.000 & 2.41  & 12.51 \\
& noisy    & 93.3 & 0.209 & 3.32  & 15.57 \\
\cmidrule(lr){2-6}
& enh. & 226.4 & 0.228 & \textbf{2.88}  & \textbf{12.28} \\
& ours     & \textbf{99.4}  & \textbf{0.207} & 3.19  & 14.66 \\
\midrule
\multirow{4}{*}{\begin{tabular}{l}URGENT\\2024\end{tabular}}
& clean    & 0.0  & 0.000 & 2.36  & 10.95 \\
& noisy    & 32.1 & 0.195 & 3.33  & 15.74 \\
\cmidrule(lr){2-6}
& enh. & 79.0 & 0.216 & \textbf{0.84}  & 10.69 \\
& ours     & \textbf{69.3} & \textbf{0.205} & 0.82  & \textbf{11.08} \\
\midrule
\multirow{4}{*}{\begin{tabular}{l}VCTK\\DEMAND\end{tabular}}
& clean    & 0.0  & 0.000 & 3.43  & 14.60 \\
& noisy    & 8.1  & 0.147 & --    & --    \\
\cmidrule(lr){2-6}
& enh. & 3.2  & 0.099 & \textbf{0.62}  & 1.93  \\
& ours     & \textbf{3.0}  & \textbf{0.098} & \textbf{0.62}  & \textbf{1.98}  \\
\midrule
\multirow{4}{*}{\begin{tabular}{l}MSP\\SNR-4\end{tabular}}
& clean    & 0.0  & 0.000 & 0.62  & 3.01  \\
& noisy    & 21.1 & 0.163 & 0.78  & 3.02  \\
\cmidrule(lr){2-6}
& enh. & 38.9 & 0.192 & 0.69  & 2.72  \\
& ours     & \textbf{29.9} & \textbf{0.181} & \textbf{0.67}  & \textbf{2.82}  \\
\midrule
\multirow{4}{*}{\begin{tabular}{l}MSP\\SNR-8\end{tabular}}
& clean    & 0.0  & 0.000 & 0.62  & 3.01  \\
& noisy    & 16.0 & 0.136 & 0.69  & 2.76  \\
\cmidrule(lr){2-6}
& enh. & 27.1 & 0.170 & 0.65  & 2.62  \\
& ours     & \textbf{21.4} & \textbf{0.159} & \textbf{0.62}  & \textbf{2.72}  \\
\bottomrule
\end{tabular}%
} 
\end{table}

Table~\ref{tab:subjective-metrics} summarizes the subjective listening test. The average inter-annotator agreement was 54.77\%, probably due to diverse distortion conditions that increase the subjectivity and variability of perceptual ratings. The degraded performance of the initially enhanced speech compared to noisy speech suggests that listeners are less tolerant of unnatural distortions introduced by SE, as reported in~\cite{chen2022inqss}. Our method mitigates this issue by detecting unreliable segments, restoring them with the original signal, and improving listener preference over the raw enhanced speech.

\begin{table}[t]
\centering
\caption{Subject ratings on URGENT24 subset (mean {\scriptsize$\pm$} std).}
\label{tab:subjective-metrics}
\setlength{\tabcolsep}{5pt}
\renewcommand{\arraystretch}{1.00}
\resizebox{0.8\linewidth}{!}
{%
\begin{tabular}{cccc}
\toprule
\textbf{clean} & \textbf{noisy} & \textbf{enhanced} & \textbf{corrected}\\
\midrule
$4.18_{\pm 0.83}$  & $3.02_{\pm 0.97}$ & $2.30_{\pm 1.43}$  & $2.87_{\pm 1.06}$ \\
\bottomrule
\end{tabular}%
} 
\end{table}

\section{Conclusion}
We introduce a correction module applied to existing SE models to mitigate destructive over-suppression while preserving denoising benefits under unseen noise conditions. Results show that our method is compatible with various SE architectures and generally improves performance across datasets in both SA metrics and downstream evaluations compared with SE baselines. This work focuses on reconstruction quality, with computational efficiency and real-time deployment left for future investigation. Future work could also explore specialized SA models for over-suppression detection, more effective failure localization, and feedback-based refinement. Overall, our work demonstrates the advantages of decoupling the initial enhancement and correction stages and suggests potential directions for speech processing pipelines.

\bibliographystyle{IEEEbib}
\bibliography{strings,refs}

\end{document}